\documentclass[12pt,preprint]{aastex}
\usepackage{color}
\shorttitle{X-ray and Optical Excess Emission in GRBs}
\shortauthors{Asano, Inoue \& M\'esz\'aros}

\begin{document}

\title{
Prompt X-ray and Optical Excess Emission due to Hadronic Cascades in Gamma-Ray Bursts}
\author{\scshape Katsuaki Asano\altaffilmark{1},
Susumu Inoue\altaffilmark{2}, and
Peter M\'esz\'aros\altaffilmark{3,4}}
\email{asano@phys.titech.ac.jp, inoue@tap.scphys.kyoto-u.ac.jp, nnp@astro.psu.edu}

\altaffiltext{1}{Interactive Research Center of Science, 
Tokyo Institute of Technology, 2-12-1 Ookayama, Meguro-ku, Tokyo 152-8550, Japan}
\altaffiltext{2}{Department of Physics, 
Kyoto University, Oiwake-cho, Kitashirakawa, Sakyo-ku, Kyoto 606-8502, Japan}
\altaffiltext{3}{Department of Astronomy \& Astrophysics;
Department of Physics;
Center for Particle Astrophysics;
Pennsylvania State University,
University Park, PA 16802}
\altaffiltext{4}{Fermi National Accelerator Laboratory,
Batavia, IL 60510}

\date{Submitted; accepted}

\begin{abstract}
A fraction of gamma-ray bursts exhibit distinct spectral features in their prompt emission below few 10s of keV
that exceed simple extrapolations of the low-energy power-law portion of the Band spectral model.
This is also true for the prompt optical emission observed in several bursts.
Through Monte Carlo simulations,
we model such low-energy spectral excess components as hadronic cascade emission
initiated by photomeson interactions of ultra-high-energy protons accelerated
within GRB outflows.
Synchrotron radiation from the cascading, secondary electron-positron pairs
can naturally reproduce
the observed soft spectra in the X-ray band, and in some cases the optical spectra as well.
These components can be directly related to the higher energy radiation at GeV energies due
to the hadronic cascades.
Depending on the spectral shape, the total energy in protons
is required to be comparable to or appreciably larger than the observed total photon energy.
In particular, we apply our model to the excess X-ray and GeV emission of GRB 090902B,
and the bright optical emission of the ``naked-eye'' GRB 080319B.
Besides the hard GeV components detected by {\it Fermi},
such X-ray or optical spectral excesses are further potential signatures
of ultra-high-energy cosmic ray production in gamma-ray bursts.
\end{abstract}


\maketitle

\section{Introduction}
\label{sec:intro}

Gamma-ray bursts (GRBs) are potential sources of ultra-high-energy cosmic rays
\citep[UHECRs,][]{wax95,vie95,mil96}.
While the required local UHECR emissivity is
$\varepsilon_{\rm p}^2 d\dot{N}_{\rm p}/d\varepsilon_{\rm p} \simeq 0.8 \times 10^{44}\ {\rm erg\ Mpc^{-3} yr^{-1}}$
at proton energy $\varepsilon_{\rm p} \sim 10^{19}$ eV \citep{wax98,der07},
the local GRB rate may be in the range of
$0.05-1\ {\rm Gpc^{-3} yr^{-1}}$ \citep{dai06,le07,gue07}.
Assuming $p_{\rm p}=2$ for the power-law index of accelerated protons,
the necessary total isotropic-equivalent energy in protons would be
$E_{\rm p} \sim 2 \times 10^{54} - 3 \times 10^{55}$ erg
when integrated over $\varepsilon_{\rm p} \sim 10^9 - 10^{20}$ eV.
Compared to the isotropic-equivalent energy released as gamma-rays, typically $E_\gamma \sim 10^{53}$ erg,
this indicates that the accelerated protons must dominate the energy budget of GRBs
in order for them to be viable sources of UHECRs.

Although $E_{\rm p}/E_\gamma \ga 10 - 100$ in GRBs may appear physically demanding,
a number of considerations provide some justification \citep{tot98,asa09}.
First, in the popular internal shock model where the prompt MeV gamma-rays
are attributed to synchrotron radiation from electrons accelerated in shocks within the GRB outflow \citep{pir05,mes06},
the efficiency with which the electrons are energized may be limited in comparison
with the total available energy, which is the dissipated fraction of the bulk kinetic energy
conveyed predominantly by protons.
Since simple Coulomb collisions are ineffective in
transferring the energy of the protons to the electrons within a dynamical timescale,
it has been generally assumed that this can occur through some plasma instabilities
with efficiencies $\epsilon_{\rm e} \sim 0.1$-$0.5$.
However, this is by no means physically guaranteed,
so we are motivated to consider the possibility of proton-dominated GRBs with $\epsilon_{\rm e} \ll 0.1$
and explore its consequences.
Note that a high proton-to-electron ratio is not only observed in Galactic cosmic rays
and inferred in supernova remnants \citep{aha06},
but also theoretically expected, at least for nonrelativistic shocks \citep{bla94,lev96}.

The prompt emission spectra of GRBs are known to be generally well described
by the so-called Band model, consisting of a hard, low-energy power-law part,
a softer, high-energy power-law part, and a spectral peak  in between in the sub-MeV range \citep{ban93}.
However, the Large Area Telescope (LAT) onboard the Fermi Gamma-Ray Space Telescope ({\it Fermi})
recently detected an additional, hard spectral component above $\sim 0.1$ GeV
in the prompt phase of the short GRB 090510 \citep{abd09b,ack10}.
Notwithstanding alternative models \citep{gra10,der10},
if the GRB accelerated ultra-high-energy protons with isotropic-equivalent luminosity $L_{\rm p} \gtrsim 10^{55}$ erg$/$s,
synchrotron and inverse Compton (IC) emission from an electron-positron pair cascade
triggered by photopion interactions of the protons with low-energy photons \citep{boe98,gup07,asa07}
can account for this GeV emission \citep{asa09b}.

Interestingly, in this same burst, the {\it Fermi} Gamma-ray Burst Monitor (GBM) observed
a further, soft excess feature below $\sim 20$ keV,
which appears to lie on a continuation of the hard, GeV power-law.
A similar, even clearer X-ray excess, as well as a GeV excess was also reported
for the prompt emission of the long GRB 090902B \citep{abd09c}.
Evidence of extra low-energy components was also reported in about $\sim 15$ \% of BATSE bursts \citep{pre96}.
In the optical band, several GRBs have exhibited prompt optical fluxes that are brighter
than expected from simple extrapolations of their low-energy
Band spectra \citep{yos07,pan08}.
One of the most impressive cases was the extremely luminous optical emission
of the ``naked-eye'' GRB 080319B \citep{rac08}.

Despite alternative explanations such as an early onset of the afterglow \citep{ghis09,kum09}
or upscattering of external/photospheric photons \citep{tom09,tom10,pee10},
the emission from hadronic pair cascades
can also potentially account for the excess components in the X-ray or even optical band,
owing to the generally very wide distribution in energy of the secondary pairs and their resultant radiation.
In addition to high-energy gamma-rays,
prompt X-ray and optical emission may thus prove to be
valuable observational signatures of UHECR acceleration in GRBs.
Here we demonstrate this through Monte Carlo spectral modeling,
focusing on the two remarkable cases of GRB 090902B and GRB 080319B.

\section{Model and Methods}
\label{sec:model}

Our Monte Carlo code self-consistently calculates the photon and neutrino spectra
corresponding to individual pulses in the GRB prompt emission
in a one-zone approximation, including all relevant leptonic and hadronic processes \citep{asa09b}.
More detailed descriptions of the code can be found in a series of previous publications \citep{asa05,asa06,asa07,asa09}.
In this work, we choose not to explicitly model the MeV-range Band spectral component,
whose true origin is currently under debate and may or may not be related to electrons accelerated in internal shocks.
Even in the case that photospheric emission or some other mechanism is more relevant,
it is plausible that coexisting internal shocks \citep{mes00,tom10,pee10} or magnetic reconnection regions \citep{gia10}
accelerate protons to ultra-high energies within the GRB outflow.
Here we simply assume that a radiation field consistent with the observed Band spectrum
is present within the emission region,
which also contains magnetic fields with energy density $U_B$,
and is moving out with bulk Lorentz factor $\Gamma$ at radius $R$ from the central engine.
Protons are injected in this environment with isotropic-equivalent luminosity $L_{\rm p}$
in the form of a power-law energy distribution with a high-energy cutoff,
$\propto \varepsilon_{\rm p}^{-2} \exp{(-\varepsilon_{\rm p}/\varepsilon_{\rm p,max})}$,
above $\varepsilon_{\rm p,min}=10$ GeV in the comoving frame.
The maximum proton energy $\varepsilon_{\rm p,max}$ is determined by the balance between
the acceleration timescale $t_{\rm acc}=\xi R_{\rm L}/c$, where $R_{\rm L}$ is the proton gyroradius,
and either the comoving expansion timescale $t_{\rm exp}=R/(c\Gamma)$
or the proton cooling timescale $t_{\rm cool}$ due to synchrotron and photomeson production losses.
We choose $\xi=1$ as required for UHECR production by GRBs,
and observationally inferred for supernova remnant shocks \citep{uch07}.
Of the free model parameters $R$, $\Gamma$, $U_B$ and $L_{\rm p}$,
the latter two can be quantified relative to
the photon energy density $U_\gamma=L_\gamma/(4 \pi c R^2 \Gamma^2)$ of the Band component
with isotropic-equivalent luminosity $L_\gamma$
as $U_B/U_\gamma$ and $U_{\rm p}/U_\gamma$ where $U_{\rm p}=L_{\rm p}/(4 \pi c R^2 \Gamma^2)$.

The physical processes taken into account are
1) photon emission and energy losses by synchrotron radiation and Compton scattering including the Klein-Nishina regime
for electrons/positrons, protons, pions and muons,
2) synchrotron self-absorption for electrons/positrons,
3) $\gamma \gamma$ pair production, 4) photomeson production for protons and neutrons,
5) Bethe-Heitler pair production ($p + \gamma \to p + e^+ + e^-$), and
6) decays of pions and muons.

For the results below, we do not include the effects of intergalactic $\gamma\gamma$ absorption
due to the extragalactic background light, since the latest observational constraints \citep[][and references therein]{abd10}
indicate that it should not be severe at the energies and redshifts of our interest here.

\section{Results}
\label{sec:results}

\subsection{GRB 090902B}
\label{sec:0909}

\begin{figure}[htb!]
\centering
\epsscale{1.0}
\plotone{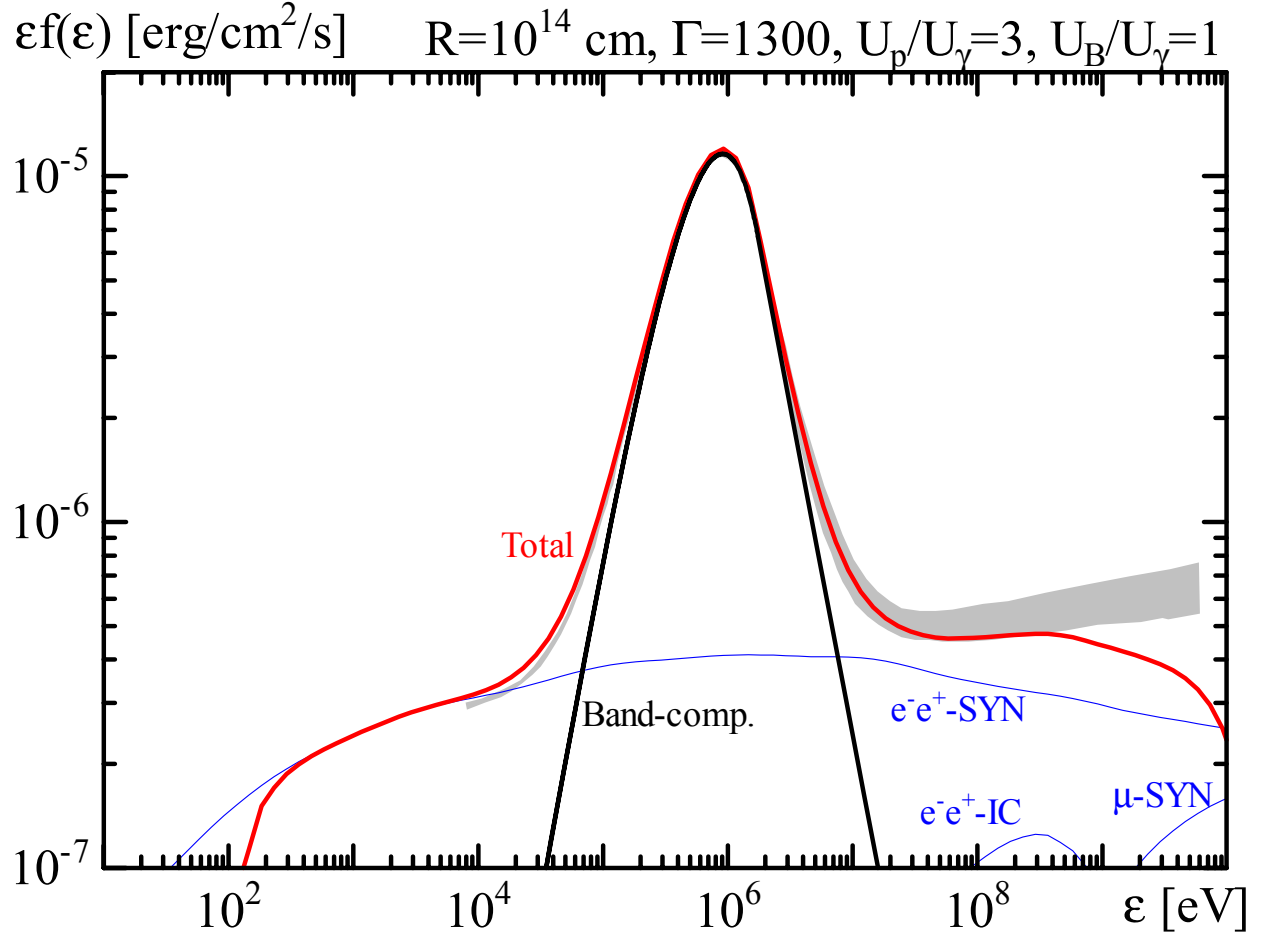}
\caption{Model spectrum for parameters listed at the top as thick red curve
compared with observations of GRB 090902B,
for which the gray shaded area represents the 1-$\sigma$ confidence region
of the best-fit, unfolded spectrum for the joint {\it Fermi} GBM and LAT data.
The best-fit Band component is shown separately as the solid black curve.
Individual contributions of synchrotron and inverse Compton from secondary electron-positron pairs
as well as muon synchrotron are denoted by thin blue curves as labelled,
not including the effects of $\gamma\gamma$ absorption or synchrotron self-absorption.
\label{fig:0909}}
\end{figure}

GRB 090902B at $z=1.822$ was one of the most energetic long GRBs
with isotropic energy $E_{\rm iso}=3.6 \times 10^{54}$ erg \citep{abd09c}.
The detection by {\it Fermi} of an 11 GeV photon together with variability on timescales $\sim 53$ ms
during the prompt phase implies a minimum bulk Lorentz factor $\Gamma_{\rm min} \sim 1000$.
Between $T_0+4.6$ s and $T_0+9.6$ s where $T_0$ is the {\it Fermi} GBM trigger time,
a hard power-law component with photon index $\Gamma_\gamma=-1.94$ was found above 0.1 GeV, 
in addition to a component well fit by a Band function with low-energy photon index $\alpha=0.07$,
high-energy photon index $\beta=-3.9$ and peak energy $E_{\rm peak}=908$ keV.
This Band component manifests an atypically narrow energy distribution, as apparent in Figure \ref{fig:0909}.
Also observed was a soft excess feature below 50 keV that is consistent with an extrapolation
of the GeV power-law down to these energies.
The gray shaded area in Figure \ref{fig:0909} represents the 1-$\sigma$ confidence region
of the best-fit, unfolded spectrum for GRB 090902B,
based on a likelihood analysis of the combined GBM and LAT data
under the assumption of a Band function plus an additional power-law \citep{abd09c}.

Overlayed in Figure \ref{fig:0909} is our model of pair cascade emission induced by ultra-high-energy protons
for the parameters $R=10^{14}$ cm, $\Gamma=1300$, $U_{\rm p}/U_\gamma=3$, and $U_{\rm B}/U_\gamma=1$.
The overall agreement is good, except for some deviation above $\sim$ GeV.
Considering the low photon statistics at the highest energies for the LAT,
and the fact that the reference spectrum is an unfolded one wherein a simple functional form was assumed,
we do not consider this to be a serious discrepancy.
Because the GeV component is characterized by photon index $\sim -2$
with relatively low power compared to the Band component,
it can be reproduced reasonably well by pair cascade emission that is dominated by synchrotron processes
in a relatively strong magnetic field \citep[e.g.][]{cop92}.
The necessary nonthermal proton luminosity is then not excessive and only comparable to the Band component luminosity.
These inferences are in notable contrast to the case of GRB 090510,
whose GeV emission was harder with photon index $\sim -1.6$ and more luminous,
warranting a stronger inverse Compton contribution from the secondary pairs
and consequently a weaker magnetic field.
This entailed a lower $\varepsilon_{\rm p,max}$ and hence lower photopion production efficiency,
leading to the requirement that $L_{\rm p}>10^{55}$ erg $\mbox{s}^{-1}$ \citep{asa09b}.

The cascade emission also extends to low energies and naturally explains the X-ray excess.
Note that a proton synchrotron model would have difficulty
accounting for such spectra with roughly equal power across a very broad energy range.
On the other hand, for the present parameters, we do not expect strong emission down to the optical band,
as the spectrum turns over at $\lesssim 100$ eV due to synchrotron self-absorption.
Moreover, the narrow energy distribution of the Band component and the corresponding $\gamma\gamma$ opacity
precludes the injection of electrons/positrons with energies low enough to radiate in the optical
(see thin blue curve in Figure \ref{fig:0909} where self-absorption is neglected).
We also mention that the high-energy cutoff due to $\gamma\gamma$ absorption expected at $\gtrsim 10$ GeV 
is a diagnostic feature to be tested in future observations of similar bursts.

\subsection{GRB 080319B}
\label{sec:0803}

GRB 080319 at $z=0.937$ was also an energetic long GRB
with isotropic gamma-ray energy $E_{\rm iso}=1.3 \times 10^{54}$ erg.
Most remarkably, it was accompanied by an extremely bright, prompt optical burst
reaching 5.3 apparent magnitude, earning the moniker ``naked-eye GRB''.
Unlike most previous prompt optical detections that could be reasonably interpreted
as emission from an external reverse or forward shock \citep{fox06,rom06,yos07},
the unprecedentedly well-sampled light curve of this burst revealed several peaks over the first $\sim 50$ s
that were correlated with the variability in MeV gamma-rays \citep{rac08,sta09,bes10},
before switching over to a behavior more typical of afterglows.
Furthermore, the early optical fluxes lie far above low-energy extrapolations
of the concurrent, time-resolved Band spectra, clearly pointing to
a distinct spectral component associated with the prompt phase \citep{rac08}.
Among different possibilities,
an interpretation attributing the optical and MeV emission
respectively to the synchrotron and synchrotron-self-Compton (SSC) processes
have often been invoked \citep[e.g.][]{rac08}.

\begin{figure}[htb!]
\centering
\epsscale{1.0}
\plotone{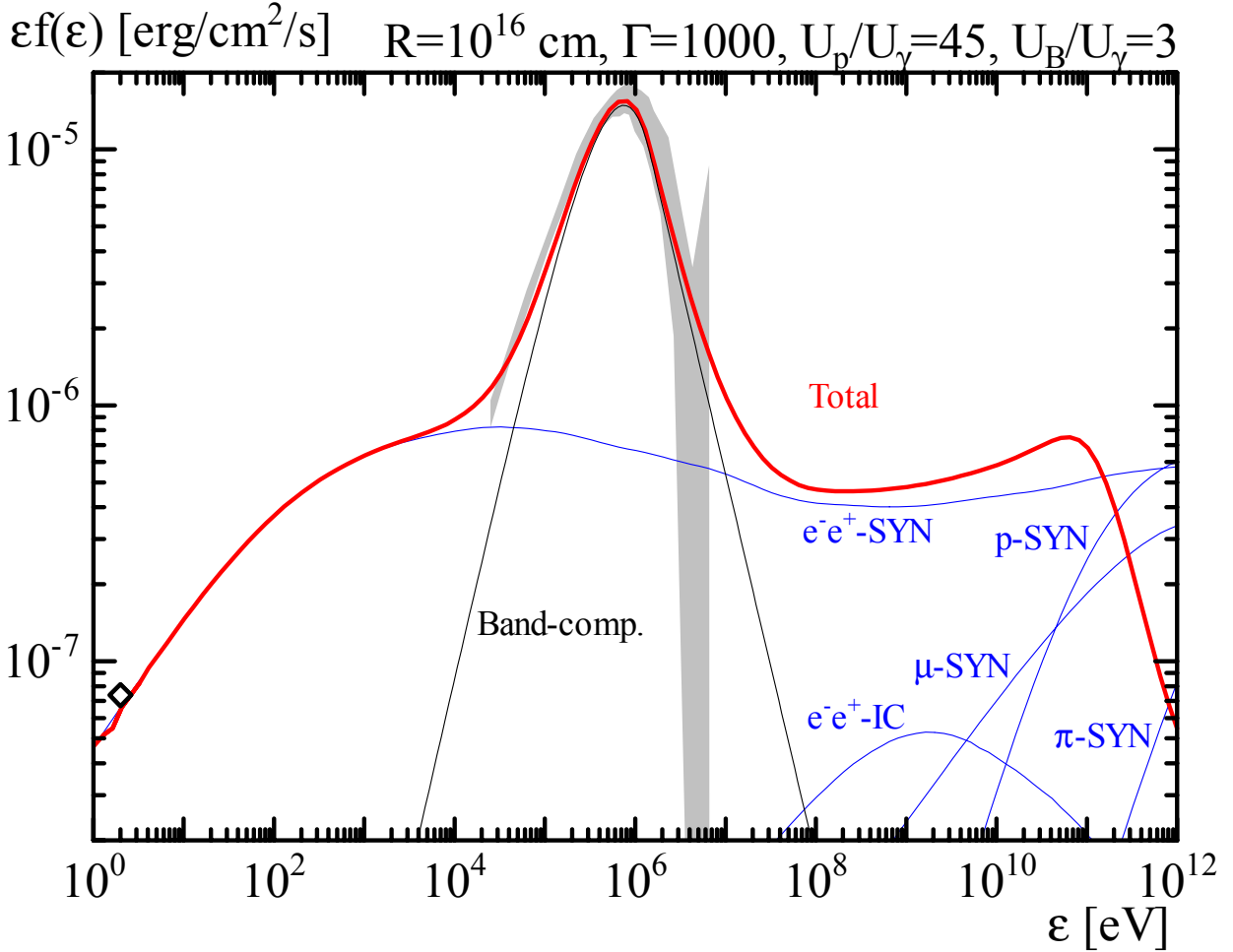}
\caption{
Model spectrum for parameters listed at the top as thick red curve
compared with observations of GRB 080319B,
for which the gray shaded area represents the spectrum measured
between $T_0+12$ s and $T_0+22$s by Swift/BAT and Konus-Wind.
The contemporaneous optical flux observed by ``Pi of the Sky'' is the black diamond.
The best-fit Band component is shown separately as the thin black curve.
Individual contributions of synchrotron and inverse Compton from secondary electron-positron pairs,
as well as muon synchrotron and proton synchrotron are denoted by thin blue curves as labelled,
not including the effects of $\gamma\gamma$ absorption or synchrotron self-absorption.
\label{fig:0803}}
\end{figure}

For comparison with our hadronic cascade model, we focus on the time interval
between $T_0+12$ s and $T_0+22$s where $T_0$ is the Swift/BAT trigger time,
during which the Band component reaches a maximum luminosity
of $1.0 \times 10^{53}~\mbox{erg}$ $\mbox{s}^{-1}$.
The Band parameters we adopt are
$\alpha=-0.45$, $\beta=-3.5$, and $E_{\rm peak}=748$ keV,
noting that $\beta$ is only constrained by the data to be $\beta<-3.412$.
Figure \ref{fig:0803} illustrates our results
for the parameters $R=10^{16}$ cm, $\Gamma=1000$, $U_{\rm p}/U_\gamma=45$, and $U_{\rm B}/U_\gamma=3$.

Synchrotron emission from the pair cascade softens the spectrum below $\sim 100$ keV
so that the low-energy slope of the Band component
becomes concordant with the observed value of $\alpha=-0.816$.
The cascade emission continues down into the optical band with some curvature
and accounts well for the observed optical intensity
as long as synchrotron self-absorption does not set in,
which necessitates $\Gamma>1000$ and $R>10^{16}$ cm.
Such large values for $\Gamma$ and $R$ imply a relatively low comoving photon density
and thus low photopion production efficiency, which scales as $\propto R^{-1} \Gamma^{-2}$.
This in turn calls for a proton luminosity $L_{\rm p} \sim 10^{55}~\mbox{erg} \mbox{s}^{-1}$,
strongly dominating that in the Band component.
Note that a substantially larger energy budget than in MeV gamma-rays alone
is also unavoidable in the SSC model for GRB 080319B
due to luminous, second-order IC emission in the GeV band \citep{rac08}.

Occurring before the launch of {\it Fermi},
the GeV properties of GRB 080319B remains largely unknown.
Upper limits at the level of $\sim 10^{-5} \ \mbox{erg}~\mbox{cm}^{-2}~\mbox{s}^{-1}$
above $10$ GeV were obtained by MILAGRO \citep{mil10}.
This does not contradict the high-energy component expected in our model,
consisting of synchrotron emission from secondary pairs
with luminosity $\sim 0.1 L_\gamma$ and photon index $\sim -2$,
extending up to a cutoff $\sim 100$ GeV 
due to internal $\gamma \gamma$-absorption (Figure \ref{fig:0803}).
The spectral shape is conspicuously different from the sharply peaked one
expected from second-order IC emission in the SSC interpretation \citep{rac08},
providing an important distinguishing feature for UHECR-induced emission components.
The overall similarity of our model spectrum for GRB 080913B with that of GRB 090902B 
also encourages us to search for optical signatures in future {\it Fermi} GRBs.

\section{Conclusions and Discussion}
\label{sec:conc}

Emission from pair cascades initiated by photopion production of ultra-high-energy protons in GRB outflows
can yield spectra with roughly equal power over a broad energy range.
While GeV-TeV signatures due to hadronic processes have been discussed previously by many authors \citep[e.g.][]{der10},
the relevance of associated X-ray or optical features had not received much attention.
Here we showed that synchrotron emission from hadronic cascades can reproduce
the excess X-ray and GeV components in GRB 090902B,
as well as the bright optical emission in GRB 080319B.
Unlike the case of GRB 090510, the necessary proton luminosity for GRB 090902B
is only comparable to the observed photon luminosity.
On the other hand, GRB 080319B calls for an isotropic-equivalent proton luminosity 
$L_{\rm p} \sim 10^{55}~\mbox{erg}$ $\mbox{s}^{-1}$,
which may appear extreme but is in fact consistent with the energetics requirements
for GRBs to be the origin of UHECRs \citep{asa09}.

The detailed time variability properties of the low-energy excess components
are not yet known observationally, except for the well-documented optical light curve of GRB 080319B
\citep{rac08,sta09,bes10} \citep[see also][]{ves04,bla04}.
Although an in-depth discussion is beyond the scope of this paper,
we may expect the variability of hadronic cascade emission to be
influenced by the timescale of photopion production 
and thus qualitatively different from purely leptonic processes.
There should also be a close connection between
the excess components at low and high energies.
The X-ray excess may be better characterized in the near future
by the Joint Astrophysics Nascent Universe Satellite (JANUS),
which will conduct prompt GRB observations including the 1-20 keV band.
The continuing development of wide-field and/or rapidly-slewing telescopes \citep{mun10}
should bring forth a clearer picture in the optical band for more bursts.
The broadband variability of hadronic emission components will be discussed
at greater length in subsequent work.

In order to explain the complex behavior of the afterglow of GRB 080319B,
a two-component jet model was discussed in \citet{rac08},
with a narrower and faster jet dominating the early phase of the afterglow,
and a wider and slower jet describing the later phase.
Such composite models would loosen the constraints
that we obtained here for our one-zone model. 
If the wider jet is the main site of proton acceleration,
and if it is dense enough to cause $p-p$ collisions in the prompt phase,
there may be other parameter sets that can reproduce the optical emission. 
Furthermore, the jet collimation-corrected energy in protons can be diminished
in such a composite model.

One of the problems for the internal shock synchrotron model of the prompt emission
is the difficulty in accounting for the observed low-energy spectra of the Band component,
which are generally much harder than naive expectations with $\alpha=-1.5$,
including the GRBs dealt with in this paper.
Many authors have proposed modifications to the model
\citep[e.g.][and references therein]{asa09c}
or alternatives such as those based on photospheric emission
\citep[e.g.][]{mes00}. 
Although here we have tacitly assumed the presence of the Band component as is observed,
in the future we intend to model all spectral components self-consistently
within the context of physical processes
that can reconcile the low-energy index observations,
in order to achieve a more comprehensive understanding of emission mechanisms
together with UHECR production in GRBs.

\begin{acknowledgments}
This study is partially supported by Grants-in-Aid for Scientific Research
No.22740117, No.22540278, and No.19047004 from the Ministry of Education,
Culture, Sports, Science and Technology (MEXT) of Japan,
and NASA NNX09AT72G, NASA NNX08AL40G, and NSF PHY-0757155.
Also acknowledged are U.R.A. and
Grants-in-Aid  for the global COE program
{\it The Next Generation of Physics, Spun from Universality and Emergence}
at Kyoto University
from MEXT.
\end{acknowledgments}


\begin{thebibliography}{}

\bibitem[Abdo et al. (2009b)]{abd09b}
Abdo, A. A. et al. 2009, \nat, 462, 331
\bibitem[Abdo et al. (2009c)]{abd09c}
Abdo, A. A. et al. 2009, \apj, 706 L138
\bibitem[Abdo et al. (2010)]{abd10}
Abdo, A. A. et al. 2010, arXiv:1005.0996
\bibitem[Ackermann et al. (2010)]{ack10}
Ackermann, M. et al. 2010, \apj, 716, 1178
\bibitem[Aharonian et al. (2006)]{aha06}
Aharonian, F. A. et al. 2006, \aap, 449, 223
\bibitem[Asano (2005)]{asa05}
Asano, K. 2005, \apj, 623, 967
\bibitem[Asano \& Nagataki (2006)]{asa06}
Asano, K., \& Nagataki, S. 2006, \apjl, 640, L9
\bibitem[Asano et al. (2009b)]{asa09b}
Asano, K., Guiriec, S., \& M\'esz\'aros, P. 2009, \apj, 705  L191
\bibitem[Asano \& Inoue (2007)]{asa07}
Asano, K., \& Inoue, S. 2007, \apj, 671, 645
\bibitem[Asano et al. (2009)]{asa09}
Asano, K., Inoue, S., \& M\'esz\'aros, P. 2009, \apj, 699, 953
\bibitem[Asano \& Terasawa (2009)]{asa09c}
Asano, K., \& Terasawa, S. 2009, \apj, 705, 1714
\bibitem[Aune (2010)]{mil10}
Aune, T. 2010, proceedings of the conference "The Shocking Universe", Venice, Italy, 2009
\bibitem[Band et al. (1993)]{ban93}
Band, D. et al. 1993, \apj, 413, 281
\bibitem[Beskin et al. (2010)]{bes10}
Beskin, G. et al. 2010, \apj, 719, L10
\bibitem[Blake et al. (2004)]{bla04}
Blake, C. H. et al. 2004, \nat, 435, 181
\bibitem[Blandford (1994)]{bla94}
Blandford, R. D. 1994, \apjs, 90, 515
\bibitem[B\"ottcher \& Dermer (1998)]{boe98}
B\"ottcher, M., \& Dermer, C. D. 1998, \apj, 499, L131
\bibitem[Coppi (1992)]{cop92}
Coppi, P. S. 1992, \mnras, 258, 657
\bibitem[Daigne et al. (2006)]{dai06}
Daigne, F., Rossi, E., \& Mochkovitch, R. 2006, \mnras, 372, 1034
\bibitem[Dermer (2007)]{der07}
Dermer, C. D. 2007, arXiv:0711.2804
\bibitem[Dermer (2010)]{der10}
Dermer, C. D. 2010, arXiv:1008.0854
\bibitem[Fox \& Meszaros (2006)]{fox06}
Fox, D. B., \& Meszaros, P. 2006, New J. Phys., 8, 199
\bibitem[Ghisellini et al. (2009)]{ghis09}
Ghisellini, G. at al. 2010, \mnras, 403, 926
\bibitem[Giannios (2010)]{gia10}
Giannios, D., arXiv:1007.1522
\bibitem[Granot (2010)]{gra10}
Granot, J. arXiv:1003.2452
\bibitem[Guetta \& Piran (2007)]{gue07}
Guetta, D., \& Piran, T. S. 2007, JCAP, 07, 003
\bibitem[Gupta \& Zhang (2007)]{gup07}
Gupta, N., \& Zhang, B. 2007, \mnras, 380, 78
\bibitem[Kumar \& Duran (2009)]{kum09}
Kumar, P., \& Duran, R.B., arXiv:0910.5726.
\bibitem[Le \& Dermer (2007)]{le07}
Le, T., \& Dermer, C. D. 2007, \apj, 661, 394
\bibitem[Levinson (1996)]{lev96}
Levinson, A. 1996, \mnras, 278, 1018
\bibitem[M\'esz\'aros (2006)]{mes06}
M\'esz\'aros, P. 2006, Rep. Prog. Phys., 69, 2259
\bibitem[M\'esz\'aros \& Rees (2000)]{mes00}
M\'esz\'aros, P., \& Rees, M. J. 2000, \apj, 530, 292
\bibitem[Milgrom \& Usov (1996)]{mil96}
Milgrom, M. \& Usov, V. 1996, Astropart. Phys., 4, 365
\bibitem[Mundell et al. (2010)]{mun10}
Mundell, C. G. et al. arXiv:1003.3573
\bibitem[Panaitescu (2008)]{pan08}
Panaitescu, A. arXiv:0811.1235
\bibitem[Pe'er et al. (2010)]{pee10}
Pe'er, A. et al. arXiv:1007.2228
\bibitem[Preece et al. (1996)]{pre96}
Preece, R. et al. 1996, \apj, 473, 310
\bibitem[Piran (2005)]{pir05}
Piran, T. 2005, Rev. Mod. Phys., 76, 1143
\bibitem[Racusin et al. (2008)]{rac08}
Racusin, J. L. et al., 2008, \nat, 455, 183
\bibitem[Roming et al. (2006)]{rom06}
Roming, P. et al. 2006, \apj, 652, 1416
\bibitem[Stamatikos et al. (2009)]{sta09}
Stamatikos, M. et al., arXiv:0902.0263
\bibitem[Toma et al. (2009)]{tom09}
Toma, K., Wu, X.-F. \& M\'esz\'aros, P. 2009, \apj, 707, 1404
\bibitem[Toma et al. (2010)]{tom10}
Toma, K., Wu, X.-F. \& M\'esz\'aros, P. 2010, arXiv:1002.2634
\bibitem[Totani (1998)]{tot98}
Totani, T. 1998, \apj, 509, L81
\bibitem[Uchiyama et al. (2007)]{uch07}
Uchiyama, Y. et al. 2007, \nat, 449, 576
\bibitem[Vestrand et al. (2004)]{ves04}
Vestrand, W. T. et al. 2004, \nat, 435, 178
\bibitem[Vietri (1995)]{vie95}
Vietri, M. 1995, \apj, 453, 883
\bibitem[Waxman (1995)]{wax95}
Waxman, E. 1995, \prl, 75, 386
\bibitem[Waxman \& Bahcall (1998)]{wax98}
Waxman, E. \& Bahcall, J. 1998, \prd, 59, 023002
\bibitem[Yost et al. (2007)]{yos07}
Yost, S. et al. 2007, \apj, 657, 925

\end{thebibliography}
\end{document}